\documentclass[epj]{webofc}
\usepackage[varg]{txfonts}   

\usepackage{units}
\usepackage{siunitx}
\usepackage{upgreek}

\usepackage{xcolor}
\definecolor{darkred}{rgb}{0.5,0,0}
\definecolor{darkblue}{rgb}{0,0,0.5}
\definecolor{firebrick}{rgb}{0.75,0.125,0.125}
\definecolor{darkgreen}{rgb}{0,0.5,0}
\usepackage[colorlinks=true,linkcolor=firebrick,citecolor=darkgreen,urlcolor=darkblue]{hyperref}

\renewcommand*{\eqref}[1]{Eq.~(\ref{eq:#1})}

\def\vvB{\vec{v} \times  (\vec{v} \times \vec{B})}
\def\xmax{X$_\mathrm{max}$\xspace}

\wocname{ARENA-2016}

\woctitle{ARENA-2016}

\begin{document}

\title{Simulation of the Radiation Energy Release in Air Showers}

\author{\firstname{Christian} \lastname{Glaser}\inst{1}\fnsep\thanks{\email{glaser@physik.rwth-aachen.de}} \and \firstname{Martin} \lastname{Erdmann}\inst{1} \and \firstname{J\"org R.} \lastname{H\"orandel}\inst{2,3} \and \firstname{Tim} \lastname{Huege}\inst{4} \and \firstname{Johannes} \lastname{Schulz}\inst{2}}

\institute{RWTH Aachen University, III. Physikalisches Institut A, Aachen, Germany
\and
           IMAPP, Radboud University Nijmegen, Nijmegen, Netherlands
\and
           Nikhef, Science Park, Amsterdam, Netherlands
\and
	  Karlsruhe Institute of Technology, Institut für Kernphysik, Karlsruhe, Germany
          }

\abstract{
  A simulation study of the energy released by extensive air showers in the form of MHz radiation is performed using the CoREAS simulation code. We develop an efficient method to extract this radiation energy from air-shower simulations. We determine the longitudinal profile of the radiation energy release and compare it to the longitudinal profile of the energy deposit by the electromagnetic component of the air shower. We find that the radiation energy corrected for the geometric dependence of the geomagnetic emission scales quadratically with the energy in the electromagnetic component of the air shower with a second order dependency on the atmospheric density at the position of the maximum of the shower development $X_\mathrm{max}$. In a measurement where $X_\mathrm{max}$ is not accessible, this second order dependence can be approximated using the zenith angle of the incoming direction of the air shower with only a minor deterioration in accuracy. This method results in an intrinsic uncertainty of 4\% with respect to the electromagnetic shower energy which is well below current experimental uncertainties.
}

\maketitle

\section{Introduction}
\label{intro}
The measurement of high-energy cosmic rays using short radio pulses emitted by air showers is a quickly evolving field of research \cite{Huege2016}. Recently, a new method to measure the cosmic-ray energy using the radiation energy, i.e., the energy that is emitted by the air shower within the frequency band of the detector, was presented \cite{ICRC2015CGlaser, AERAEnergyPRD, AERAEnergyPRL, ARENA2016GlaserAERA}. In this work, we study the emission of the radiation energy from the theoretical side using Monte Carlo simulations of air showers and the calculation of the radiation from first-principles based on classical electrodynamics. More details of this analysis can be found in \cite{GlaserErad2016a}.

\section{Method}
We use the CoREAS code \cite{CoREAS2013a}, which is an extension of the CORISKA \cite{CORSIKA} software, to simulate the radio emission from air showers. It uses a full MC approach where in principle all shower particles are followed and their contributions to the radiation at a specific observer position are calculated. We simulated 592 air showers with energies between \SI{e17}{eV} and \SI{e19}{eV}, zenith angles between 0$^\circ$ and 80$^\circ$ and uniformly distributed azimuth angles. For each combination of energy and geometry we simulate a proton- and an iron-initiated air shower. 

As the radiation at each observer position needs to be calculated separately, the radiation energy is no direct outcome of the simulation but needs to be determined via integration of the radio emission footprint. As this would require a large number of sampling points and the computing time scales linearly with the number of observers, we developed a method that exploits the radial symmetry of the geomagnetic and the charge-excess contributions in the radio signal. We found that it is sufficient to simulate the radiation only at positions along the positive $\vvB$-axis, where $\vec{v}$ is the shower direction and $\vec{B}$ is the geomagnetic field. This is because on this axis the geomagnetic and charge-excess radiation are polarized orthogonally to each other and can therefore be separated. 
This method does not only allow for an efficient extraction of the radiation energy from air-shower simulations but also allows for a separate determination of the radiation energy resulting from the geomagnetic and the charge-excess radiation. 

\section{Longitudinal Profile of the Radiation Energy Release}

\begin{figure}[t]
\centering
\sidecaption
\includegraphics[width=0.65\textwidth]{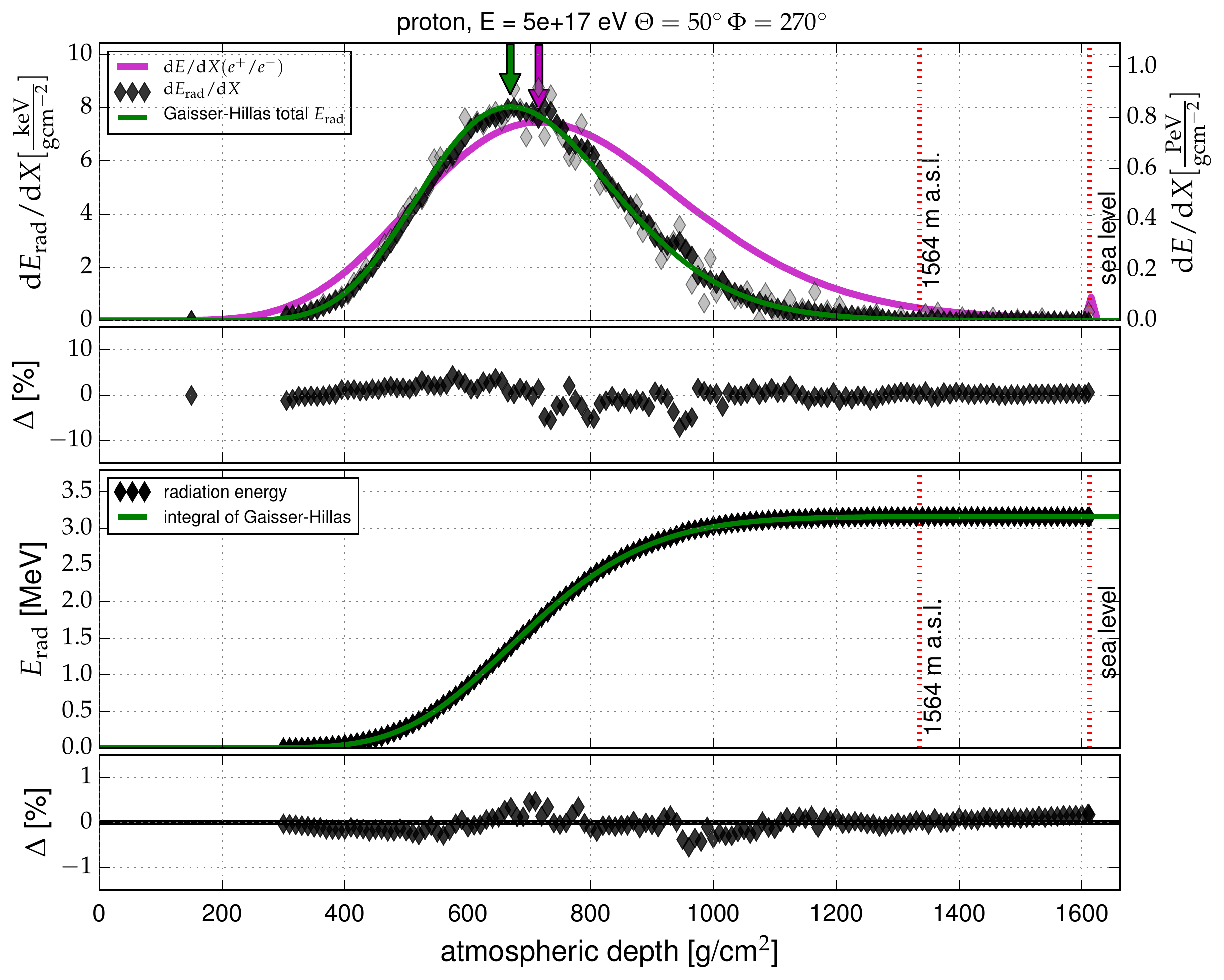}
\caption{Longitudinal profile of the radiation energy release. The lower diagram plots the total radiation energy measured at a given atmospheric depth. The green line demonstrates a fit of the integral of a Gaisser-Hillas function to the radiation energy. The residuals are also shown. The upper diagram plots the same data and fit, but differentially, i.e., as radiation energy release at a given atmospheric depth. Also shown is the energy deposit of the electromagnetic shower particles as a purple line that needs to be read on the right vertical axis. The arrows indicate the position of the shower maximum. Figure and caption adapted from \cite{GlaserErad2016a}.}
\label{fig:long}
\end{figure}

To determine at which stage of the shower development how much radiation energy is emitted, we simulate observers at different atmospheric depths along the shower axis. The resulting longitudinal profile of the radiation energy release of one of the simulated air showers is presented in Fig.~\ref{fig:long}. The radiation energy rises with increasing atmospheric depth until it reaches its maximum and then remains constant as the atmosphere is transparent to radio waves.
This does not imply that the lateral signal distribution remains constant. The signal distribution changes strongly with the distance between observation height and shower maximum. For larger distances the radiation energy is distributed over a larger area resulting in a broader signal distribution with smaller signal strengths. 

The longitudinal profile of the radiation energy release can be described with a Gaisser-Hillas function with three free parameters. We compared the profile of the radiation energy release with the longitudinal profile of the energy deposit of the air shower ($\mathrm{d}E/\mathrm{d}X$) and found that the radio profile is shifted to smaller atmospheric depths with respect to the $\mathrm{d}E/\mathrm{d}X$ profile with an average shift of \SI{46}{g/cm^2}.

\section{Correlation with Electromagnetic Shower Energy}
To achieve a good correlation of the radiation energy with the electromagnetic shower energy, the dependence of the geomagnetic emission on the angle $\upalpha$ between the geomagnetic field and the shower axis needs to be corrected for. As the charge-excess emission is independent of the geomagnetic field we only correct for the geomagnetic contribution to the radiation energy by dividing the radiation energy by $[a^2 + (1-a^2) \sin^2\upalpha]$, where $a$ is defined as the square root of the ratio of the charge-excess and the geomagnetic contribution to the radiation energy normalized to maximum geomagnetic emission ($\sin \upalpha = 1$). We use the square root of the ratio to be consistent with previous work on this topic where $a$ was defined as the ratio of electric-field amplitudes \cite{AERAPolarization, LofarPolarization2014}. 

With our method, the geomagnetic and charge-excess contributions to the radiation energy can be extracted separately from the simulated air showers. We found that $a$ is not constant but depends on the atmospheric density in which the shower develops. The larger the atmospheric density in which the shower develops the larger the ratio of charge-excess to geomagnetic radiation energy. We observe values ranging from $a$ = \SI{0.02} to $a$ = 0.5. 
This effect can be parametrized with the density at the shower maximum $\rho_{X_\mathrm{max}}$ which is primarily a function of the zenith angle and depends to a smaller extend on the variations of the shower maximum \xmax. This dependence is described by the following exponential function:
\begin{equation}
 a(\rho_\mathrm{X_\mathrm{max}}) = 0.43  \left(e^{\unit[1.11]{m^3/kg} \, (\rho_\mathrm{X_\mathrm{max}} - \langle \rho \rangle)}\right) - 0.24\, ,
 \label{eq:adxmax}
\end{equation}
where we use $\langle \rho \rangle$ = \unit[0.65]{kg/m$^3$} as the average atmospheric density at the shower maximum.

In addition to the $\sin \upalpha$ dependence we found that the radiation energy itself also depends on the atmospheric density in which the shower develops. For shower developments in small density regions we observe an increased radiation energy. This is because the shower development depends on the amount of atmosphere that is traversed (slant depth) whereas the amount of radiation depends on the geometric path length of the shower particles. For showers developing in thin regions of the atmosphere the ratio of geometric path length to slant depth is larger than for showers developing in thick regions resulting in an increased radiation energy. 

\begin{figure}[t]
\centering
\sidecaption
\includegraphics[width=0.7\textwidth]{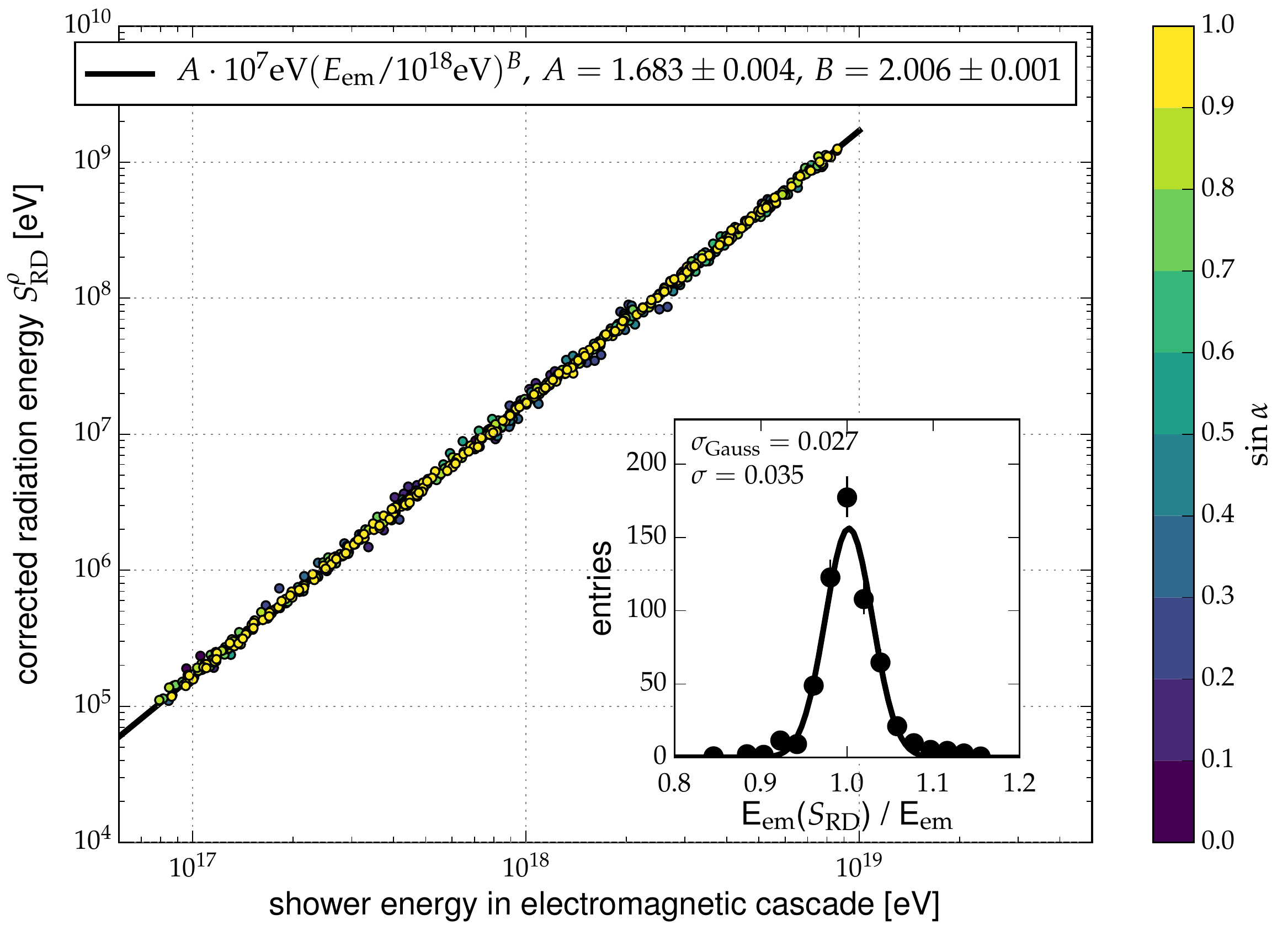}
\caption{Correlation between the energy in the electromagnetic part of the air shower and the corrected radiation energy (cf. Eq.~(\ref{eq:Srddxmax})). The black line shows a power-law fit to the data. The inset figure shows the scatter around the calibration curve. The color scale shows the value of $\sin \upalpha$. Figure from \cite{GlaserErad2016a}.}
\label{fig:ecal}
\end{figure}

This dependence can again be parametrized using $\rho_{X_\mathrm{max}}$ and we add a second correction term to the radiation energy.

\begin{equation}
S_\mathrm{RD}^{\rho} = \frac{E_\mathrm{rad}}{a(\rho_\mathrm{X_\mathrm{max}})^2 + (1-a(\rho_\mathrm{X_\mathrm{max}})^2) \sin^2\upalpha} \,\, \frac{1}{\left(
1 - p_0 + p_0 \,  \exp[p_1 (\rho_{X_\mathrm{max}} - \langle \rho \rangle)]\right)^2 } \,.
 \label{eq:Srddxmax}
\end{equation}
The correlation between the corrected radiation energy $S_\mathrm{RD}^{\rho}$ and the electromagnetic shower energy is presented in Fig.~\ref{fig:ecal}. We describe the correlation with a function of the form $S_\mathrm{RD} = A \times \unit[10^7]{eV} \,(E_\mathrm{em}/\unit[10^{18}]{eV})^B$ and determine the free parameters $A$ and $B$ as well as the parameters $p_0$ and $p_1$ of Eq.~(\ref{eq:Srddxmax}) in a combined $\chi^2$ fit. We find $A = 1.683$, $B = 2.006$, $p_0 = 0.251$ and $p_1 = \SI{-2.95}{m^3/kg}$. Hence, the corrected radiation energy scales quadratically with the electromagnetic shower energy as expected for coherent emission. The intrinsic energy resolution of this method can be inferred from the scatter around the calibration curve and amounts to $\sim3\%$ which is shown in the inset figure. 

In a measurement, \xmax is often not accessible or has large experimental uncertainties. We therefore approximated the density at the shower maximum using only the zenith angle of the incoming shower direction and assuming an average value of \xmax for all air showers. With this simplification we still obtain an intrinsic uncertainty of only 4\% which is well below current experimental uncertainties. 

\section{Conclusion}
In this work, we presented a prediction of the radiation energy emitted by air showers using first-principles calculations based on classical electrodynamics. We determined the longitudinal profile of the radiation energy release. Furthermore, we studied the dependence on the shower energy and found that the radiation energy scales quadratically with the energy in the electromagnetic cascade of the air shower after correcting for the dependence of the geomagnetic emission on the geomagnetic field. In addition, we found that the radiation energy shows a second-order dependence on the atmospheric density in which the shower develops. This dependence can be parametrized using the atmospheric density at the shower maximum resulting in an intrinsic uncertainty of the method of 3\%. In a more practical parametrization using only the zenith angle, the method shows an intrinsic uncertainty of 4\%. Hence, the results presented here can be used by cosmic-ray radio experiments to improve the precision in the energy reconstruction and to calibrate the energy scale from first-principles calculations.

\end{document}